\documentclass[12pt,a4paper]{article}
\usepackage{epsf,pifont}
\begin{document}

\baselineskip 20pt     %preprint

\begin{center}

{\bf
MULTIFRAGMENTATION WITH INDRA@GSI:\\
FROM THE FERMI TO THE PARTICIPANT-SPECTATOR DOMAIN}

%(draft \today)
\vspace*{0.6cm}

\normalsize               

W. Trautmann,$^{(a)}$ A. Le F{\`e}vre,$^{(a)}$ J. {\L}ukasik$^{(a)}$\\
and the INDRA and ALADIN Collaborations$^{(b)}$\\
\vspace*{0.3cm}
   {\small \it  $^{(a)}$ Gesellschaft f{\"u}r Schwerionenforschung (GSI), Planckstr. 1, D-64291 Darmstadt, Germany \\
   $^{(b)}$ Ganil Caen, LPC Caen, INFN Catania, GSI Darmstadt, INFN Napoli, IPN Lyon, IPN Orsay, DAPNIA Saclay, SINS Warsaw, IFJ Krak{\'o}w, INR Moscow, CNAM Paris \\[8mm]}

\end{center}

\begin{abstract}
With INDRA at GSI, multifragmentation was studied in $^{197}$Au + $^{197}$Au reactions at bombarding energies 40 to 150 MeV per nucleon. The mechanisms of fragment production at mid-rapidity in peripheral collisions and the interplay of collective motion and fragmentation in central collisions were investigated. The balance energy for squeeze-out, i.e. the transition energy from in-plane to out-of-plane enhancement of the particle and fragment emissions is determined as a function of the impact parameter.

\end{abstract}

\normalsize               

\section{Introduction}
\label{sec:intr}

In 1997, the 4$\pi$ INDRA multidetector~\cite{pouthas} has been brought to GSI with the aim to use the unique capabilities of this instrument for reaction studies at energies beyond those available at GANIL. In a series of experiments, conducted within three subcampaigns in 1998 and 1999, the symmetric $^{197}$Au + $^{197}$Au and Xe + Sn systems and the asymmetric $^{12}$C + $^{197}$Au and $^{12}$C + $^{112,124}$Sn reactions were studied over wide ranges of bombarding 
energies.
Isotopic effects were studied by cross bombarding isotopically pure $^{112,124}$Sn targets with beams of $^{124,129}$Xe at 100 MeV per nucleon. With the $^{12}$C beams, the decay of the produced target residues, nearly at rest in the laboratory, was investigated at incident energies 95 to 1800 MeV per nucleon~\cite{turzo04}. 

In this contribution, we concentrate on the $^{197}$Au + $^{197}$Au system, studied at 40 to 150 MeV per nucleon, a range of energies that corresponds to relative velocities between once and twice the Fermi value. The reduction of Pauli blocking and the opening of the phase space for nucleon-nucleon collisions in the entrance channel cause rapid changes in the phenomenology of the collision processes whose understanding was one of the primary goals of this campaign. In particular, the onset and development of flow phenomena in more central collisions were expected to strongly influence the fragmentation of the system~\cite{lavaud01}. Rapid changes are also observed in the more peripheral collisions; the centers of fragment production move from mid-rapidity, characteristic for the Fermi energy domain, toward the spectator rapidities as the energy is increased (Fig. 1). At 150 MeV per nucleon, the emission patterns approach those familiar from relativistic bombarding energies.

\begin{figure}[!htb]
     \epsfysize=10.0cm
   
     \centerline{\epsffile{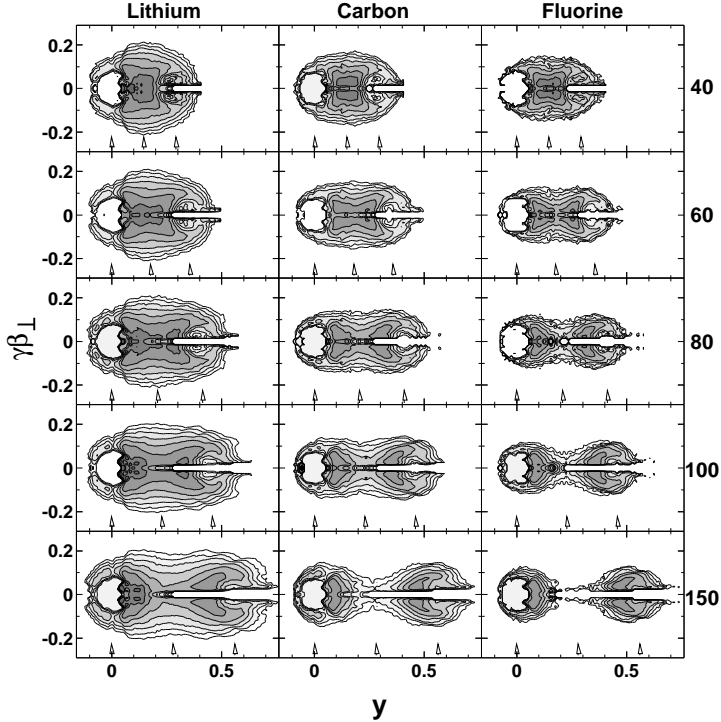}}

     \caption{\footnotesize{
     Invariant cross section distributions as a function of
     transverse velocity  $\gamma\beta_{\perp}$ and rapidity $y$ for
     fragments with $Z$ = 3, 6, 9 from the  most peripheral collisions
     (bin 1) of $^{197}$Au + $^{197}$Au at $E/A$ =  40, 60, 80, 100 and
     150 MeV, as indicated. The cross sections are normalized 
     relative to
     each other, separately for each fragment species. Near the target
     rapidity they are affected by the thresholds for fragment
     identification. The contour scale is logarithmic; the arrows 
     (from left to right) indicate the target, center-of-mass and 
     projectile rapidities (from~\protect\cite{lukasik03}).
}}
\label{fig:au15}
\end{figure}

\section{Experimental details}
\label{sec:expe}

\begin{figure}[!htb]
     \epsfxsize=7.0cm
     \centerline{\epsffile{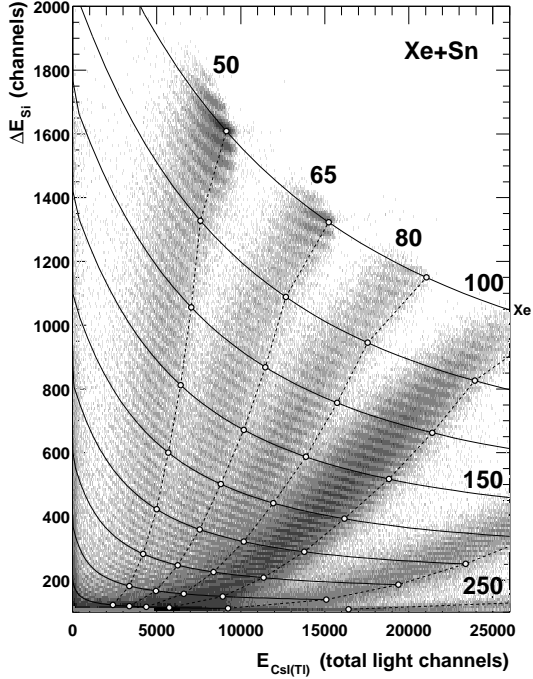}}

     \caption{\footnotesize{
     Scatter plot of $\Delta E$ versus $E$ signals derived from the 
     300-$\mu$m Si detectors and the 14-cm CsI(Tl) detectors of a 
     module of ring 1 for fragments from
     Xe + Sn reactions at several energies between 50 and 250 MeV per 
     nucleon. The grid represents lines of equal atomic number $Z$ 
     (solid lines) and equal kinetic energy per nucleon (dashed) and 
     is obtained from energy-loss and range tables using the 
     calibration adopted for this detector.
}}
\label{fig:xesn}
\end{figure}

Technical details of the INDRA@GSI experiments, including the calibration and identification procedures, have been described in Refs.~[2-7].
%\cite{turzo04,lavaud01,lukasik03,lukasik02,trzcinski03,lefevre04}. 
As an example, the identification map for a module of ring 1 (2$^{\circ} \leq \theta_{\rm lab} \leq 3^{\circ}$), from the Xe + Sn subcampaign is shown in Fig. \ref{fig:xesn}. For the measurements at GSI, this first ring had been modified in order to improve the energy and charge resolution. Without changes of the ring geometry, the previously used phoswich detectors were replaced by 12 telescopes, each consisting of a 300-$\mu$m Si detector and a CsI(Tl) scintillator viewed by a photomultiplier. This is the same technique that is also used in the neighbouring INDRA rings up to $\theta_{\rm lab} \leq 45^{\circ}$. The figure shows that the fragment atomic numbers are individually resolved up to $Z$ = 54 of the xenon beam. The quasielastic fragment groups were used to check the energy calibration.

For the impact parameter selection, the total transverse energy $E^{12}_{\perp}$ of light charged particles ($Z$ = 1,2) was chosen to serve as the sorting variable for most analyses. 
Eight impact parameter bins were used, with the most central bin 8 covering impact parameters up to 5\% of $b_{\rm max}$. The remaining part of the $E^{12}_{\perp}$ spectrum was then divided into 7 bins of equal width, corresponding to 7 bins of  approximately equal width in $b$. Bin 1 contains the most peripheral events (cf. Fig. \ref{fig:au15}).

\section{Fragment formation in peripheral collisions}
\label{sec:peri}

\begin{figure}[!htb]
     \epsfysize=10.0cm
     \centerline{\epsffile{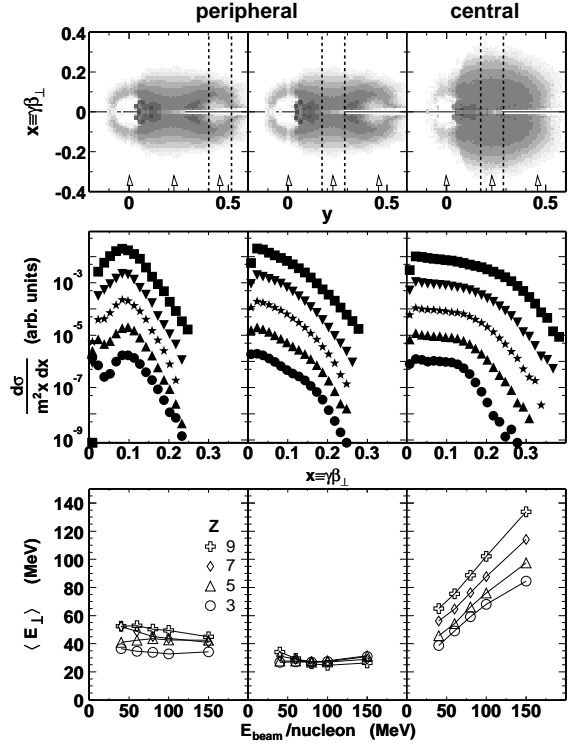}}

     \caption{\footnotesize{
     Top row: invariant cross section distributions for $Z$ = 3 
     fragments as a function of transverse velocity $\gamma\beta_{\perp}$ and 
     rapidity $y$ for peripheral (bin 1, left and center) and 
     central (bins 7 and 8, right) collisions of $^{197}$Au + $^{197}$Au 
     at $E/A$ = 100 MeV. The dashed lines indicate the windows in 
     relative rapidity chosen for the data shown below; the arrows 
     denote the rapidities of the target, the center-of-mass, and 
     the incident projectile (from left to right).
     \newline
     Middle row:
     invariant transverse velocity spectra
     for $Z$ = 3 at bombarding energies $E/A$ = 40, 60, 80, 100 
     and 150 MeV (from bottom to top and vertically displaced for 
     clarity).
     \newline
     Bottom row: mean transverse energies $\langle E_{\perp} \rangle$
     as a function of the incident energy for fragments with odd
     $Z \leq$ 9 as indicated (from~\protect\cite{lukasik02}).
}}
\label{fig:main}
\end{figure}

Important insight into the mechanism of fragment production in peripheral collisions is gained from the study of the transverse fragment motion as, e.g., represented by invariant transverse-velocity spectra (Fig. \ref{fig:main}). These spectra are expected to be Gaussian for a thermally emitting source,
with a width $\sigma^2 = T/m$ where $T$ and $m$ are the temperature of the 
source and the mass of the emitted particle, respectively. If
Coulomb forces act in addition, a peak will appear near the velocity
corresponding to the Coulomb energy.

The examples of transverse-velocity 
spectra of lithium nuclei given in the middle row of panels in Fig. \ref{fig:main} represent three distinct cases: emissions at projectile rapidity in peripheral collisions (left panel), mid-rapidity and peripheral collisions (middle) and mid-rapidity emission in central collisions (right). The selected rapidity intervals are 
indicated in the invariant cross section plots shown in the upper row of the figure. The corresponding mean transverse energies as functions of the incident energy for fragments from lithium to fluorine are shown below.

The transverse velocity spectra at projectile rapidity in peripheral 
collisions are dominated by a prominent 
Coulomb peak that corresponds to the repulsion from
the surface of a heavy residue. The spectra are virtually the same at all bombarding energies which is reflected by the invariant mean kinetic energies (Fig. \ref{fig:main}, left panels).
Coulomb peaks are absent in the emissions at midrapidity which exhibit two
different scaling behaviors for the central and peripheral impact 
parameters. In the central case (right panels), 
the shapes are approximately Gaussian, with an extra shoulder superimposed at the lower bombarding energies, and with a width that increases rapidly with the bombarding energy. The mean transverse energies, correspondingly, exhibit a nearly linear rise as a function of the bombarding energy and increase also with the fragment mass.
These observations reflect the increasing collectivity
of the fragment motion as the incident energy rises, a result of higher 
compression, a resulting stronger Coulomb acceleration, 
and higher temperatures 
of the composite sources that are initially formed 
in central collisions~\cite{lavaud01,lefevre04,nebau99}. We will come back to this topic in the next section.

\begin{figure}[!htb]
     \epsfxsize=8.0cm
     \centerline{\epsffile{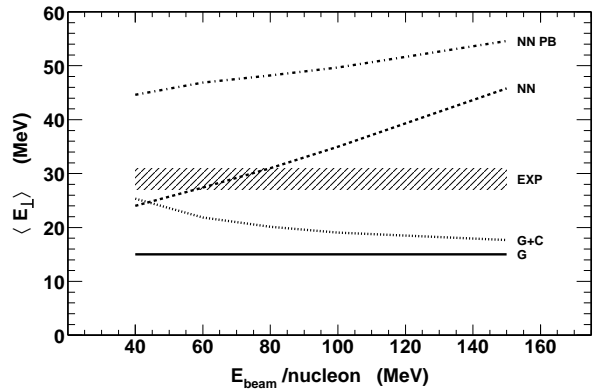}}

     \caption{\footnotesize{
     Mean transverse energies $\langle E_{\perp} \rangle$ as obtained from the
     Goldhaber model for $p_{\rm F}$ = 265 MeV/c (full line labelled G),
     after adding the Coulomb energy (dotted line, G+C), and
     for nucleons from primary nucleon-nucleon collisions with 
     (dashed-dotted line, NN PB) and without (dashed line, NN) 
     considering Pauli blocking. The experimental result for 
     midrapidity fragments from peripheral collisions is represented 
     by the hatched area labelled EXP (from~\protect\cite{lukasik02}).
}}
\label{fig:last}
\end{figure}

The most striking behavior is observed for the mid-rapidity fragments 
from peripheral reactions (middle column of Fig. \ref{fig:main}).
The shapes of the transverse-velocity spectra, somewhat between Gaussian and
exponential, show hardly any change with the incident energy. The mean transverse energies of $\langle E_{\perp} \rangle \approx$ 30 MeV are the same for all energies and fragment $Z$, even beyond the displayed range of elements from lithium to fluorine.

Kinetic energies that are independent of the particle species are expected 
within thermal models. In the present case, however, the corresponding temperature $T \approx$ 30 MeV, seems rather large and clearly exceeds the temperature range at which larger fragments can be expected to survive. Kinetic energies
that appear thermal and correspond to high temperatures are also obtained from
the Goldhaber model in which
fragment momenta are assumed to result from the nucleonic Fermi
motion~\cite{gold74}. This approach has proven useful for the
interpretation of kinetic energies of intermediate-mass fragments from
spectator decays at relativistic bombarding energies~\cite{odeh00}. In the present case, however, the value
$\langle E_{\perp} \rangle$ = $T$ = 15 MeV, derived for the Fermi
momentum $p_{\rm F}$ = 265 MeV/c of heavy nuclei, amounts to only about
half of the equivalent temperature that is required (Fig. \ref{fig:last}).

A summary of other possible contributions to the transverse fragment velocities is given in Fig. \ref{fig:last}. The additional Coulomb repulsion by the projectile and target residues, derived from three-body Coulomb trajectory calculations, is not sufficient to explain the observations. The transverse energies generated in primary nucleon-nucleon collisions, on the other hand, are large, in particular, if the Pauli blocking of final states with low transverse momentum at the lower incident energies is taken into account. The figure thus suggests that one or several hard scattered nucleons will have to be incorporated into a fragment in order to raise its transverse energy up to the level that is observed. The lower Coulomb contribution at higher incident energies, when the residue velocities are becoming comparable to or larger than those of the mid-rapidity fragments, may be expected to be compensated by the larger scattering contributions in order to preserve the observed invariance.

\begin{figure}[!htb]
     \epsfxsize=9.0cm

     \centerline{\epsffile{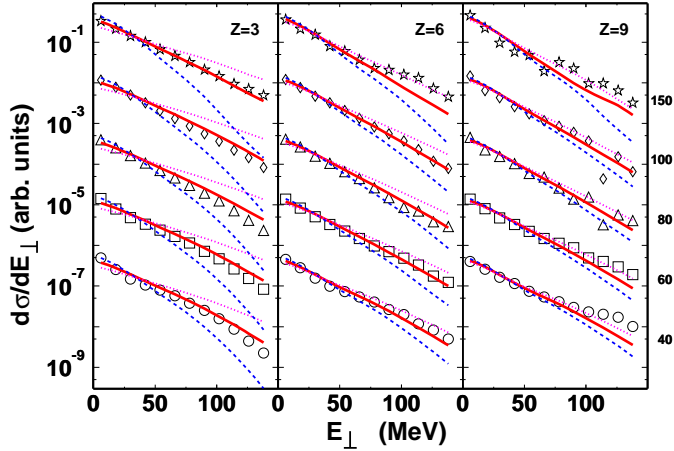}}

     \caption{\footnotesize{
     Experimental (symbols) and calculated (lines) transverse-energy
     spectra of $^7$Li (left panel), $^{13}$C (middle), and 
     $^{19}$F fragments 
     (right) emitted at mid-rapidity in 
     peripheral $^{197}$Au + $^{197}$Au
     collisions (bin 1) at five incident energies, as indicated. 
     Solid, dashed and
     dotted lines represent the calculations with on average one, 
     zero or two scattered nucleons in a fragment, respectively. 
     In each panel, the experimental spectra are displaced vertically 
     by consecutive factors of 30, and the calculated spectra are 
     individually normalized relative to the corresponding measurement
     (from~\protect\cite{lukasik03}).
}}
\label{fig:ene}      

\end{figure}

This scenario has been tested with a Monte-Carlo procedure intended to permit more quantitative predictions~\cite{lukasik03}. The original idea of Goldhaber~\cite{gold74} of taking the fragment momentum as the sum of the momenta of the
constituent nucleons picked randomly from a single Fermi sphere has been
extended by including also the Pauli-allowed distribution of hard scattered
nucleons. The calculations proceed in three main steps: 
(i) the momenta of the $A$ nucleons of a fragment with mass number $A$  are
randomly picked from 2 Fermi spheres ($p_{\rm F}$ = 265 MeV/c),
separated by a relative momentum per
nucleon calculated at the distance of closest approach of the two nuclei
colliding at a given impact parameter, and also from the momentum distribution
generated in hard two-body collisions of nucleons from the projectile and
target;
(ii) a clustering criterion has been implemented, and only fragments with internal distributions of nucleon momenta compatible with that of a single Fermi sphere with radius $p_{\rm F}$ = 265 MeV/c are retained; (iii) accepted fragments are placed, in coordinate space, in between the
two  residues in an aligned sticking configuration which serves as the
starting point for 3-body Coulomb trajectory calculations.

The resulting spectra agree quantitatively with those measured if, for the very peripheral events, on average one scattered nucleon from the Pauli-allowed distribution of hard scattered nucleons is included in the fragment
(Fig.~\ref{fig:ene}, solid lines). They are also invariant with bombarding energy and $Z$. The harder spectra at smaller impact parameters require correspondingly larger numbers of hard-scattered nucleons. 
With this prescription, also the exponentially decreasing $Z$ distributions and, in particular, the characteristic change of their slopes with the incident energy are reproduced on a quantitative level~\cite{lukasik03}.

These results support earlier findings that the mid-rapidity fragment emission cannot be considered as a superposition of yields from equilibrated target and projectile residues~[11-13].
%\cite{montoya94,lukasik97,plagnol00}. 
The calculated cross section distributions, furthermore, extend into the projectile and target rapidity regions, as a consequence of the combined action of the initial momenta and of the Coulomb forces, and exhibit the characteristic Coulomb holes (cf. Fig. \ref{fig:au15}). Apparently, also fragments observed further away from mid-rapidity do not necessarily have to be considered as
emitted from the excited residues but may to a large part originate
from the contact zone, or neck region~\cite{baran04}, formed during the reaction.

\section{Statistical fragmentation and flow in central collisions}
\label{sec:cent}

In central collisions, fragment emission is nearly isotropic in velocity space, suggesting the formation of a single hot source (Fig. \ref{fig:main}, right panels). The collective velocity fields, on the other hand, show that global equilibrium is not reached. The question to what extent the formation and survival of fragments are possible in this explosive environment has been raised and discussed by many authors~[15-18].
%\cite{kunde95,pal,chikazumi,das}. 
Of specific interest is whether statistical models can still be applied. 

\begin{figure}

  \epsfxsize=8.0 cm

     \centerline{\epsffile{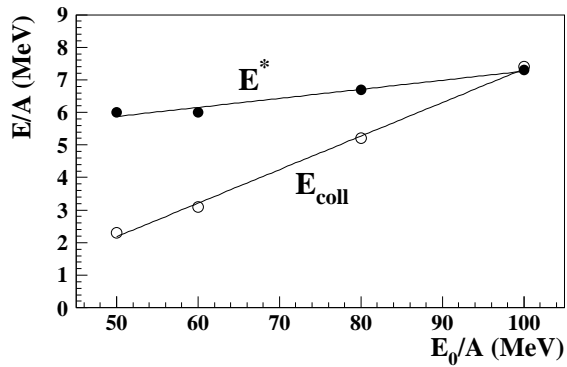}}

    \caption{\footnotesize{
    Mean thermal excitation energy
    (full circles) and collective flow energy (open circles)  at
    freeze-out, extracted by means of the MMMC-NS model,
    as a function of the incident energy $E_0/A$ for
    central collisions of $^{129}$Xe + $^{\rm nat}$Sn at 50 and 
    $^{197}$Au + $^{197}$Au at 60,
    80 and  100 MeV per nucleon. The lines are linear fits, meant to
    guide the eye (from~\protect\cite{lefevre04}).
  }}
  \label{fig:syn}
\end{figure}

Studies conducted with the Copenhagen~\cite{bond95} and Berlin versions~\cite{gross90} of the Statistical Multifragmentation Model and based on the present data have shown that the partition into fragments are well described if appropriate global parameters are used~\cite{lavaud01,lefevre04}. For the reproduction of the fragment kinetic energies, however, the addition of collective velocity components is both mandatory and sufficient. The emerging picture is summarized in Fig. \ref{fig:syn} which represents the dependence on bombarding energy of the thermal and collective energy components of the fragmenting source. The required thermal excitation energy rises slowly (by $\approx20\%$)
while the collective energy increases rapidly, about
twice as fast as the available center-of-mass energy.
The two quantities become approximately equal at a beam energy of 100 MeV per nucleon. A further important global parameter is the mass (or charge) parameter entering the statistical description. It decreases from about 90\% of the total system mass at 40 MeV per nucleon~\cite{lavaud01} to only 60\% at 100 MeV per nucleon and reflects the rapidly rising pre-equilibrium component that is emitted prior to the breakup into fragments.  The evolving fireball, with an intensity of 40\% at 100 MeV per nucleon, is linked to hard nucleon-nucleon collisions in the entrance channel and becomes the dominant reaction component as the incident energy is further increased into the relativistic regime~\cite{reis97}.

\begin{figure}

  \epsfxsize=11.0 cm

     \centerline{\epsffile{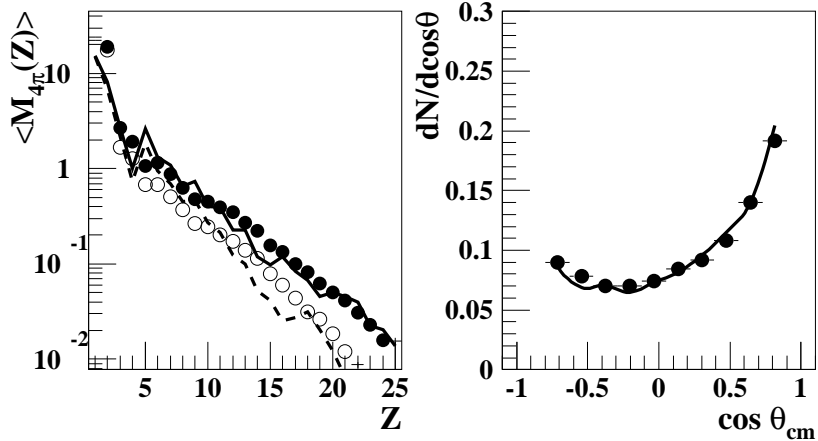}}

    \caption{\footnotesize{
    Left: Measured differential fragment multiplicity d$M$/d$Z$, 
    normalized to
    a solid angle of $4\pi$, for central $^{197}$Au + $^{197}$Au collisions at 
    60 MeV per nucleon (circles) and MMMC-NS model predictions 
    calculated for a
    prolate source with axis ratio 0.7:1.
    The full (open) circles and the solid (dashed) line correspond to 
    forward (sideward) emission angles.
    \newline
    Right: Angular distribution of the largest fragment 
    within an event for the same reaction.
    The full circles and the solid line represent
    the experimental data and the MMMC-NS model predictions,
    respectively.
    The forward-backward
    asymmetry is caused by the limited experimental acceptance
    for heavy fragments near target rapidity
    (from~\protect\cite{lefevre04}).
  }}
  \label{fig:auz}
\end{figure}

The observed small anisotropies of the fragment source, evident in the $Z$ distributions and in the angular distributions of the largest fragments (Fig. \ref{fig:auz}) as well as in the kinetic energy spectra, have been analyzed with the MMMC-NS version~\cite{lefevre99} of the Statistical Multifragmentation Model which allows for non-spherical sources~\cite{lefevre04}. The preferential emission of heavier fragments in forward/backward directions is accounted for by assuming a longitudinally elongated source with axis ratio 0.7:1. In the model, the anisotropy is the result of correlations between the charge of a fragment and its location in the freeze-out configuration, created by the mutual Coulomb interactions inside the non-spherical source. The anisotropy of the kinetic energies is reproduced with anisotropic flow profiles of similar elongation 0.7:1.  

The origin of these anisotropies is of particular interest. They are most likely related to the mechanisms of stopping and energy dissipation during the initial stages of the reaction. Incomplete stopping and a corresponding transparency of the collision partners for each other may be revealed by applying an isospin tracer method~\cite{rami} to the data from the cross bombardment of the Xe + Sn system at 100 MeV per nucleon. The existence of a prolate deformation in coordinate space has been confirmed by the fragment-fragment velocity correlations deduced from the measured data for the same reaction~\cite{lefevre03}. 

These results do not directly resolve the question of how the fragmentation mechanism is modified by the collective motion. In fact, with the possibility of a partial transparency even in the most central collisions, the successes of the Statistical Fragmentation Models is even more intriguing. Within a quantum statistical model, it has been demonstrated that the effect of the flow on the charge distributions may be simulated by simply increasing the value of the thermalized energy in the model description~\cite{pal}. This would mean that local equilibrium is sufficient for a statistical description and that the flow effect is implicitly included in the parameters. At moderate flow values, the changes are expected to be very small~\cite{das}. 

An alternative approach to the question of the coexistence of equilibrated partitions and collective motion has recently been presented by Campi \textit{et al.}~\cite{campi}. 
Using classical molecular dynamics calculations and specific clustering
algorithms, these authors find fragments to be preformed at the beginning of
the expansion stage when the temperature and density are still high.
The fragment charge distributions, reflecting the equilibrium at this early
stage when the flow is small, remain nearly unmodified down to the
freeze-out density at which the flow has fully developed. High chemical temperatures~\cite{serfling} might be possible indicators for this scenario of early fragment formation~[27-30].
%\cite{dani92,dorso95,barz96,puri96}.

\begin{figure}[hbt]

  \epsfxsize=12.0cm
     \centerline{\epsffile{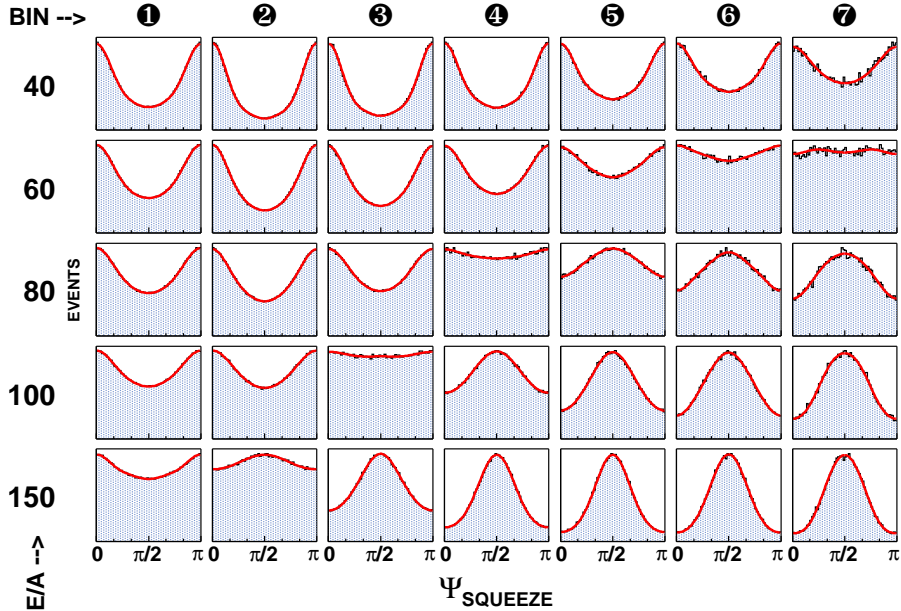}}

     \caption{\footnotesize{
     Distributions of the squeeze angle $\Psi$ for $^{197}$Au + $^{197}$Au 
     collisions as a function of the 
     bombarding energy $E/A$ and sorted into seven bins of increasing 
     centrality (from left to right). 
}}
  \label{fig:psi}
\end{figure}

\section{Squeeze-out balance}
\label{sec:sque}

A rich phenomenology of flow patterns emerges if the fragment motion is studied in its relation to the reaction plane~\cite{reisritt97}. The lowest orders of the azimuthal anisotropies are denoted as directed in-plane flow and elliptic flow, the latter describing the predominance of either in-plane or out-of-plane emissions. At the present energies, the sign of the directed flow~\cite{lemmon,cussol02} and the orientation of the elliptic flow both depend on the relative strengths of the mean-field dynamics with respect to the collision dynamics resulting from hard nucleon-nucleon scatterings. This competition causes a disappearance and reappearance of the collective velocity components at the so-called transition energies. For the elliptic flow, the transition from in-plane to out-of-plane enhancement (squeeze-out) in $^{197}$Au + $^{197}$Au was known to occur below or around 100 MeV per nucleon~\cite{andronic01} and thus within the energy range covered with the INDRA@GSI campaign. A second transition from squeeze-out back to in-plane flow at ultrarelativistic energies has recently been identified~\cite{magestro}. 

The squeeze angle $\Psi$ is defined as the angle of the second-largest of the three eigenvectors of the kinetic-energy tensor with the reaction plane~\cite{gutbrod90}. It requires the reconstruction of the reaction plane of an event for which several methods have been developed \cite{reisritt97}. In the present case, it was approximated by the plane defined by the largest eigenvector of the kinetic-energy tensor and the beam axis. The $\Psi$-distributions obtained for $^{197}$Au + $^{197}$Au collisions at five bombarding energies and for seven impact-parameter bins are shown in Fig. \ref{fig:psi}. Maxima at angles 0 and $\pi$ indicate preferential in-plane emission while maxima at $\pi$/2 characterize enhanced  emissions in directions perpendicular to the reaction plane. The distributions are flat near the transition energies which depend both on the incident energy and on the centrality. For mid-peripheral collisions for which the anisotropies are largest the transition is found to occur close to 80 MeV per nucleon.

\begin{figure}[hbt]

  \epsfxsize=9.0cm
     \centerline{\epsffile{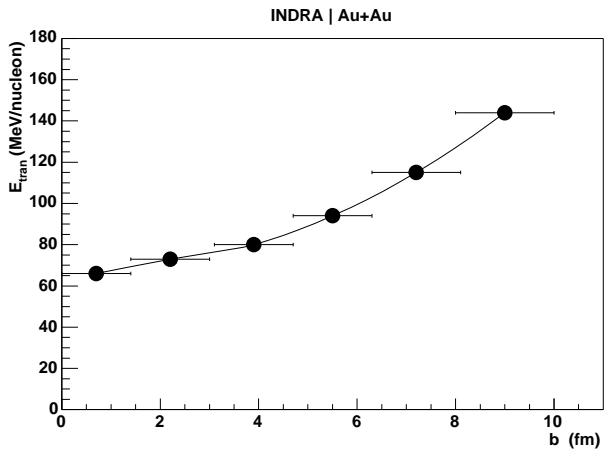}}

     \vspace{-1mm}
     \caption{\footnotesize{
     Transition energy from in-plane to out-of-plane enhancement 
     for $Z$ = 2 particles from $^{197}$Au + $^{197}$Au collisions as a 
     function of the impact parameter $b$.
}}
  \label{fig:bal}
\end{figure}

The transition energies, also called balance energies, are particularly useful for theoretical interpretations. The prediction of a balance requires the cancellation of the contributing momentum components and may thus be sensitive to specific parameters of the theoretical description.
A wealth of transition energies can be obtained if the azimuthal distributions of specific types or groups of particles as a function of their transverse momenta are considered individually. The first two coefficients from a Fourier analysis, usually denoted as $v_1$ and $v_2$, permit a quantitative characterization of the directed and elliptic flow~\cite{andronic01,gutbrod90}. The transition from predominantly attractive to repulsive in-plane flow is associated with a minimum of the slope $\partial v_1/ \partial y$ of the $v_1$ coefficient at mid-rapidity. The transition from in-plane to out-of-plane enhancement is observed as a sign change of $v_2$. As an example, the squeeze-out balance obtained for $Z$ = 2 particles, integrated over the covered range of transverse momentum, is shown in Fig.~\ref{fig:bal}. At larger impact parameters, the transition occurs at higher incident energies. Plotted as a function of the fragment charge $Z$ and for a fixed impact parameter range, the transition energies for squeeze-out tend to slowly decrease with increasing $Z$. Larger anisotropies are generally observed if higher transverse momenta are selected.

\section{Conclusion and outlook}
\label{sec:conc}

The energy range of 40 to 150 MeV per nucleon of the present study of the $^{197}$Au + $^{197}$Au system is characterized by several qualitative transitions. Ultimately, they are all connected to the reduction of Pauli blocking and the increasing importance of direct nucleon-nucleon collisions in the entrance-channel dynamics. Evidence for their role in the fragment formation in peripheral collisions is obtained from the large transverse velocities of fragments formed at mid-rapidity. Mid-rapidity is also where the fragment emission has its maximum at the lower bombarding energies. As the relative velocity of projectile and target increases beyond the Fermi value, a transition toward preferential emission near the residue rapidities is caused by the clustering condition. The limit it represents for the internal fragment momenta suppresses a mixing of unscattered nucleons from the collision partners.

Central collisions are characterized by the onset and development of the collective radial motion. At the same time, the developing fireball of preequilibrium particles contains a rapidly increasing fraction of the system. 
Correspondingly, the mass of the fragmenting source accessible to equilibrium descriptions decreases. Its parameters as obtained from the analysis with the Statistical Fragmentation Model reflect, beyond that, very little of the increasing overall violence of the collision. The rise of the mean thermal excitation energy is slow. 

The flow can be treated independently in this model description by superimposing a collective velocity field. Its influence on the partitions is, apparently, either small or remains within what can be accomodated by adjusting some of the global parameters. This somewhat unsatisfactory situation is avoided if a scenario of early fragment formation during the high-density phase is considered. It is supported by classical molecular dynamics calculations used to model the nuclear and Coulomb interactions during this early reaction phase. 

The growing importance of collective phenomena in the present energy range provides both a challenge and an opportunity for testing microscopic dynamical descriptions of reaction processes above the Fermi energy domain. The obtained flow data with the observed sign changes of the directed and elliptic flow components should allow for precise tests of specific model parameters. 

Stimulating discussions with A. Andronic and W. Reisdorf are gratefully acknowledged. This work has been supported by the European Community under Contract No. ERBFMG\-ECT\-950083.

\end{document}